\documentclass[letterpaper]{article}
\pdfoutput=1
\usepackage{jheppub}
\usepackage{amsmath}
\usepackage{graphicx}
\usepackage[config=altsf]{subfig}
\usepackage{feynmf}
\usepackage{array}
\usepackage{comment}
\usepackage{slashed}
\usepackage{afterpage}
\usepackage{appendix}
\usepackage{multirow}

\newcommand{\GeV}{\,\text{GeV}}
\newcommand{\MeV}{\,\text{MeV}}

\newcommand{\eV}{\,\text{eV}}

\def\GeV{\,\mathrm{GeV}}
\def\TeV{\text{TeV}}

\title{
Factoring the Strong CP Problem
}
\author[a]{Prateek Agrawal}
\affiliation[a]{Department of Physics, Harvard University, Cambridge, MA 02138, USA}
\author[b]{and Kiel Howe}
\affiliation[b]{Fermi National Accelerator Laboratory, Batavia IL 60510,
USA}

\preprint{FERMILAB-PUB-17-403-T}

\abstract{
We present a new mechanism to solve the strong CP problem using $N\geq
2$ axions, each dynamically relaxing part of the $\bar\theta$
parameter. At high energies $M\gg\Lambda_{QCD}$ the $SU(3)_{c}$ group becomes the diagonal
subgroup of an $SU(3)^{N}$ gauge group, and the non-perturbative effects in each individual $SU(3)$ factor generate a potential for the corresponding axion. The vacuum is naturally aligned to ensure $\bar\theta=0$ at low energies, and the masses of these axions can be much larger than for the standard QCD axion. This mechanism avoids the introduction of a discrete $Z_2$ symmetry
and associated `mirror' copies of the SM fermions, and also avoids the
introduction and stabilization of new light colored states to modify
the running of the QCD gauge coupling found in other heavy axion
models. This strengthens the motivation for axion-like particles
solving the strong CP problem at points beyond the standard QCD axion
curve in the $(m_a, f_a)$ plane.
}

\preprint{\today}

\begin{document}
\maketitle

\section{Introduction}
The Standard Model (SM) describes physics remarkably well at the
smallest scales probed so far. With the discovery of the Higgs boson,
the SM can be consistently extrapolated to very high energies.
However, there are
a number of puzzles in the SM that remain.
The strong CP problem is one such long standing problem. 
The QCD Lagrangian 
\begin{align}
  \mathcal{L}
  &=
  -\frac14 G^a_{\mu\nu} G^{a,\mu\nu}
  -\frac{g^2 \theta}{32\pi^2} G^a_{\mu\nu} \tilde{G}^{a,\mu\nu}
  + \bar{q} M q
\end{align}
in general violates CP, with the CP violation encoded by
\begin{align}
  \bar{\theta}
  &=
  \theta + \arg\det M
  \,.
\end{align}
Given that CP is badly broken by the weak interactions, we expect
$\bar{\theta} \sim 1$. However, 
the QCD $\theta$-angle is constrained to be very small, 
$\bar{\theta} \lesssim 10^{-10}$, from limits on the electric dipole
moment of the
neutron/Hg~\cite{Baker:2006ts,2016PhRvL.116p1601G,Afach:2015sja}.

The solutions to the strong CP problem can be classified roughly
into two categories.
One class of solutions imposes discrete symmetries like
CP as in the Nelson-Barr
mechanism~\cite{Nelson:1983zb,Barr:1984qx,Babu:1989rb,Hiller:2001qg} 
(see~\cite{Dine:2015jga} for a recent discussion)
or parity (P)~\cite{Babu:1989rb, Barr:1991qx} that forbid the
$\theta$-angle. In Nelson-Barr models there are generally no
low-energy states, while parity models often require new light colored
particles \cite{DAgnolo:2015uqq}.

The second class of solutions invokes a global $U(1)$ Peccei-Quinn
symmetry~\cite{Peccei:1977np, Peccei:1977hh}. This $U(1)_{PQ}$ has a
mixed anomaly with QCD, such that QCD non-perturbative dynamics break
PQ explicitly and the $\theta$-angle is dynamically relaxed to zero.
The most economical version of such a solution is the massless (up-)
quark solution~\cite{tHooft:1976rip}, in which case the $\theta$-angle
becomes unphysical (or equivalently is relaxed dynamically by the
$\eta'$ meson).  However, lattice calculations indicate that all
quarks are massive~\cite{Aoki:2016frl, Dine:2014dga}, excluding the
simplest model with a massless up-quark. Another elegant model that
uses the PQ symmetry is the QCD axion~\cite{Peccei:1977np,
Peccei:1977hh,Weinberg:1977ma,Wilczek:1977pj}
(see~\cite{Olive:2016xmw} for a review).  The $U(1)_{PQ}$ is
spontaneously broken in the UV, giving rise to a light pseudo-Nambu
Goldstone boson, the axion. The axion makes the $\theta$ angle
dynamical, and has a potential with a minimum at
$\bar{\theta}=0$~\cite{Vafa:1984xg}.

In our paper we focus on this axion solution to the strong CP problem.
Given the QCD axion
decay constant  $f_a$, the mass of the axion is
determined by QCD dynamics~\cite{Weinberg:1977ma}, 
\begin{align}
  m_{a,QCD}^2
  &=
  \frac{m_u m_d}{(m_u+m_d)^2}
  \frac{f_\pi^2 m_\pi^2}{f_a^2}
  \simeq
  \frac{(75.5\MeV)^2}{f_a}
  \equiv
  \frac{\Lambda_0^2}{f_a}
  \,.\label{eq:maNormal}
\end{align}
The axion couples to the standard model through dimension-5
  operators suppressed by $f_a$, and the precise mass relationship
  equation~\eqref{eq:maNormal} has motivated many existing and future
experimental efforts (see Ref~\cite{Graham:2015ouw} for a review).
In this work, we study an extension of the QCD axion framework where
instead of a single axion relaxing the vacuum to $\bar\theta=0$, two
or more axions naturally cooperate to solve the strong CP problem.
Each of these axions can have a mass much larger than the standard QCD
axion, motivating searches for axions in the $(m_a, f_a)$ plane
outside of the QCD axion window given by equation~\eqref{eq:maNormal}.

This  multi-axion solution to the strong CP problem arises in models
that extend the low-energy $SU(3)_c$ gauge group to be the diagonal
subgroup of a parent $SU(3)\times SU(3) \times...$  product gauge
group, which is broken down to $SU(3)_c$ at some high scale $M$. All
the SM quarks are taken to be charged under a single SU(3) factor of
the parent gauge group, and we introduce an axion for each individual
SU(3) factors that independently relaxes the corresponding
$\theta$-angle to 0.
Each SU(3) factor of the parent gauge group is more strongly coupled than the diagonal $SU(3)_c$ subgroup, and therefore the contributions to the axion potentials from UV instantons near the scale $M$ can be larger than the non-perturbative potential generated at low energies from QCD.  Even for the smallest extension, $SU(3)\times SU(3) \to SU(3)_c$, we find that the two axions can  have masses 
significantly larger than for the standard QCD axion.  In a follow-up
work, Ref.~\cite{MasslessQuarkPrep}, we will describe a related family
of models where each quark generation is charged under a different
SU(3) factor, and the strong CP problem can be solved without
introducing any axions degrees of freedom in a spirit similar to the massless
up quark solution.

Other mechanisms have been proposed that may increase the mass of
  the QCD axion. The models of
  Refs.~\cite{Rubakov:1997vp,Berezhiani:2000gh,Hook:2014cda,Fukuda:2015ana,Dimopoulos:2016lvn}
  also extend $SU(3)_c$ to an $SU(3)\times SU(3)$ gauge group, but
  with QCD living in a single factor instead of the diagonal subgroup.
  In these models, a discrete $Z_2$ symmetry is introduced, requiring an
  entire mirror SM matter sector. In contrast, our mechanism does not
  involve any new matter in the $SU(3)$ factors, and the only new
  low-energy states are the axions directly responsible for relaxing
  $\bar\theta$. The models of Refs.~\cite{Holdom:1982ex,
  Holdom:1985vx,Dine:1986bg,Choi:1988sy,Flynn:1987rs,Choi:1998ep}
introduce extra colored matter at an intermediate scale so that QCD
runs back to strong coupling at a high scale $M$ where $SU(3)_c$ is
embedded in a $SU(3+n)$ gauge group. However, the introduction of new
colored matter generally leads to either new hierarchy problems for
scalars or new CP phases for fermions, spoiling the solution to the
strong CP problem. In contrast, in our mechanism no new colored states
are introduced below the scale $M$, and extra CP phases that would
feed into the UV axion potential are naturally absent.

In Sec.~\ref{sec:nonpert}, we describe in detail the structure of the model and the calculation of the non-perturbative contributions to the axion potentials at the scale $M$. In Sec.~\ref{sec:UV}, we discuss the sensitivity of the mechanism to higher dimensional operators and the connection to the electroweak hierarchy problem. In Sec.~\ref{sec:pheno}, we describe the limits on these models from axion-like particle (ALP) searches.

\section{Non-perturbative Effects in Product Gauge Groups}
\label{sec:nonpert}

\subsection{$SU(3) \times SU(3)$ Model}
We start with a simple extension of the SM where the QCD gauge group
emerges
from Higgsing a product group 
\begin{align}
SU(3)_{1}\times SU(3)_{2} \to SU(3)_c 
\,. 
\end{align}
at a scale $M \gg \TeV$.  This a similar set-up to the renormalizable
coloron models of
\cite{Bai:2010dj,Chivukula:2013xka,Chivukula:2013xka,Chivukula:2015kua,Bai:2017zhj},
although in our case the scales will be far out of reach of the LHC.
The SM quarks are
charged only under the $SU(3)_1$ gauge factor, and there are no
fermions charged under $SU(3)_2$. 
We take the theory to have two
spontaneously broken anomalous $U(1)_{PQ}$ symmetries at scales $f_i >
M$, giving an axion
$a_{1,2}$ in each $SU(3)$ sector:
\begin{align}
  \mathcal{L} 
  &= 
  -\frac14 (G_1)^a_{\mu\nu}\ (G_1)^{a,\mu\nu}
  +\frac{g^2_{s1}}{32\pi^2}
  \left( \frac{a_1}{f_1} - \theta_1 \right)
  (\widetilde{G}_1)^a_{\mu\nu}\ (G_1)^{a,\mu\nu}
  \nonumber\\&\qquad\qquad
  -\frac14 (G_2)^a_{\mu\nu}\ (G_2)^{a,\mu\nu}
  +\frac{g^2_{s2}}{32\pi^2}
  \left( \frac{a_2}{f_2} - \theta_2 \right)
  (\widetilde{G}_2)^a_{\mu\nu}\ (G_2)^{a,\mu\nu}
\end{align}
The gauge couplings $g_{s{1,2}}$ and theta angles $\theta_{1,2}$ are
independent parameters not related to each other by any symmetries.
The presence of two independent axion degrees of freedom will allow
both physical $\theta$-angles to be dynamically removed. An additional
cross-coupling term for the axions $\propto \frac{a_1}{f_1}
\tilde{G}_2 G_{2} $ can not be removed by a field redefinition. It
leads to a mass-mixing between the axions, but does not destabilize
the solution to the strong CP problem or significantly change the
phenomenology of the light axion states, so we take it to vanish for
simplicity. The UV couplings of  $a_{1,2}$ to the
electroweak topological terms $W\tilde W$ and $B \tilde B$ will only
be relevant for the axion phenomenology, and will be discussed in
Sec.~\ref{sec:pheno}.
 

The theory will match to the SM at a scale $M$ where the
$SU(3)_1 \times SU(3)_2$ gauge group is Higgsed to a diagonal
$SU(3)_{c}$. For simplicity, we take the dynamics to be a
bifundamental scalar field $\Sigma_{12} : (3, \bar 3,1)$ of $SU(3)_1
\times SU(3)_2 \times U(1)_\Sigma$  with a renormalizable scalar
potential
\begin{align}
  V_\Sigma 
  &= 
 -m_\Sigma^2 {\rm Tr}({\Sigma_{12}\Sigma_{12}^\dagger})
+\frac{\lambda}{2}[{\rm Tr}({\Sigma_{12}\Sigma_{12}^\dagger})]^2
+\frac{\kappa}{2}{\rm Tr}(\Sigma_{12}\Sigma_{12}^\dagger\Sigma_{12}\Sigma_{12}^\dagger)
\end{align}
inducing a vev  \cite{Bai:2010dj,Bai:2017zhj}
\begin{equation}
\langle \Sigma \rangle = \frac{m_\Sigma}{\sqrt{\kappa+3\lambda}} \mathbb{I}_3 \equiv \frac{f_\Sigma}{2} \mathbb{I}_3.
\end{equation}
One combination of the $SU(3)$ gauge bosons becomes massive with
$M_V^2 \equiv M^2 = (g_{s_1}^2 + g_{s_2}^2)f_\Sigma^2$, and the
unbroken $SU(3)_{QCD}$ gauge symmetry is given by
\begin{align}
  G^\mu_{(SM)} 
  &= \cos\gamma\, G^\mu_{1} + \sin\gamma\, G^\mu_{2},
  \qquad 
  \tan\gamma = g_{s_1}/g_{s_2} 
\end{align}
The additional $U(1)_\Sigma$ gauge factor forbids a trilinear term in the potential, and the corresponding gauge boson absorbs the otherwise massless singlet Goldstone mode\footnote{As an alternative to gauging the extra U(1) factor, the trilinear term $\mu {\rm Det} \Sigma_{12} + {\rm h.c}$ could be introduced in the potential to lift the singlet Goldstone mode. Note that this CP phase in the trilinear term can be rotated way into the $\Sigma$ field, and the potential dynamically prefers a CP preserving $\Sigma$ vev \cite{Bai:2017zhj}.}. 

We can integrate out the heavy degrees of freedom to match to the SM
at this scale.  The tree-level matching condition gives the couplings
of the two SU(3) factors in terms of the standard model QCD coupling
evaluated at the scale $M$,
\begin{align}
  \frac{1}{\alpha_s (\mu)}   
  &= 
  \frac{1}{\alpha_{s_1}(\mu)}
  + \frac{1}{\alpha_{s_2}(\mu)}
  ,\qquad
  \mu=M
\label{eq:GaugeMatching},
\end{align}
with $\alpha = g^2/4\pi$.  Each individual gauge factor is more
strongly coupled than the SM QCD gauge coupling, and this will lead to
enhanced non-perturbative effects compared to the SM. The SM coupling
$g_s (M)$ can be obtained with the 1-loop running
from the top pole, $\alpha_{s}(m_t) = 0.10$.

The tree-level value of the SM theta term is simply
$$\bar\theta_{SM} = \bar\theta_1 + \bar\theta_2$$
Because the flavor symmetries are the same as the SM, the loop-level thresholds to $\bar\theta$ proportional to the CKM phase are negligible~\cite{Ellis:1978hq,Dugan:1984qf}. 

Since we are integrating out the heavy degrees of freedom, we must
also consider short distance non-perturbative effects in the
individual $SU(3)_1$ and $SU(3)_2$ factors, which will generate the
shift-symmetry breaking potentials for the axions.  We will be
interested in the regime where each factor is still at relatively weak
coupling near $M$, and therefore the dilute instanton gas
approximation (DGA) gives a good approximation for the
non-perturbative effects. The effective Lagrangian generated for the
axions is  \cite{tHooft:1976snw,Callan:1977gz,Andrei:1978xg}
\begin{align}
  \mathcal{L}_a
  &=
  \Lambda_1^4 \cos \left(\frac{a_1}{f_1}  - \bar\theta_1\right) +
  \Lambda_2^4 \cos \left(\frac{a_2}{f_2} - \bar\theta_2\right) +
  \frac{g_s^2}{32\pi^2}
  \left(\left( \frac{a_1}{f_1} - \bar{\theta}_1\right) 
  +\left(\frac{a_2}{f_2} - \bar{\theta}_2\right) \right) G\tilde{G}
  \label{eq:LUV}
\end{align}
The axions still couple to the low energy QCD $\tilde{G}G$ term, and
at the scale $\Lambda_{QCD}\sim 1~\GeV$ the non-perturbative SM
contributions to the axion potentials will be generated. However,
unlike the standard axion case, the axion potential can be dominated
by the higher energy contributions to the potential with scales
$\Lambda_1$ and $\Lambda_2$ generated at the scale $M$.
Although these scales $\Lambda_1$ and $\Lambda_2$ will be suppressed
by non-perturbative and chiral suppression factors, they can still
greatly exceed the scale of the standard axion potential. 

Crucially, and as is clear from the UV PQ symmetries of the theory, the new
short-distance non-perturbative contributions to the axion potentials
are exactly aligned to remove the effective $\theta$-angle,
\begin{align}
  \bar\theta_{eff} 
  &= 
  \left\langle 
  \left(\frac{a_1}{f_1}+ \bar\theta_1\right) 
  + \left(\frac{a_2}{f_2}+ \bar\theta_2\right)
  \right\rangle = 0
  \,.
\end{align}
The low-energy vacuum is in-fact guaranteed to align with $\bar\theta=0$ by a generalization of the Vafa-Witten argument \cite{Vafa:1984xg}. This result is
very different from theories which have new \emph{perturbative}
breakings of the PQ symmetry in the UV-- for example,
higher-dimensional operators  violating the PQ symmetry give axion
potentials that are generically misaligned and would spoil the axion
solution to the strong CP problem \cite{Barr:1992qq,Kamionkowski:1992mf,Ghigna:1992iv}.

We can now estimate how the mass of the two axions is affected by
these UV non-perturbative contributions to the potential.
In the $SU(3)_2$ factor, where no colored fermions are present, the
axion potential is suppressed only by the non-perturbative instanton
action. The scale can be calculated as
\cite{tHooft:1976snw,Callan:1977gz,Andrei:1978xg}
\begin{align}
  \Lambda_2^4 
  &= \int_{\rho=0}^{\rho=1/M} 
  2 \frac{d\rho}{\rho^5} D[\alpha_{s_2}(1/\rho)] 
  \label{eq:LambdaInt2}
\end{align}
where the dimensionless instanton density depends non-perturbatively
on the running gauge coupling as
\begin{align}
  D[\alpha] 
  &= 0.1
  \left(\frac{2\pi}{\alpha}\right)^6
  e^{-\frac{2\pi}{\alpha}},
\end{align}
with $\alpha(1/\rho)$ the running coupling evaluated at the scale
$\mu=\rho^{-1}$ corresponding to the size of the instanton. Higher
order corrections to the instanton density have been calculated in
Refs.~\cite{Dine:2014dga}. In this work, we use the leading order
result, and estimate the theoretical uncertainty by varying the scale
at which the density is evaluated: $D[\alpha_s(1/\rho)]\rightarrow
(D[\alpha_s(2/\rho)],D[\alpha_s(\frac{1}{2\rho})])$.

\begin{figure}[t]
\centering
\includegraphics[width=2in]{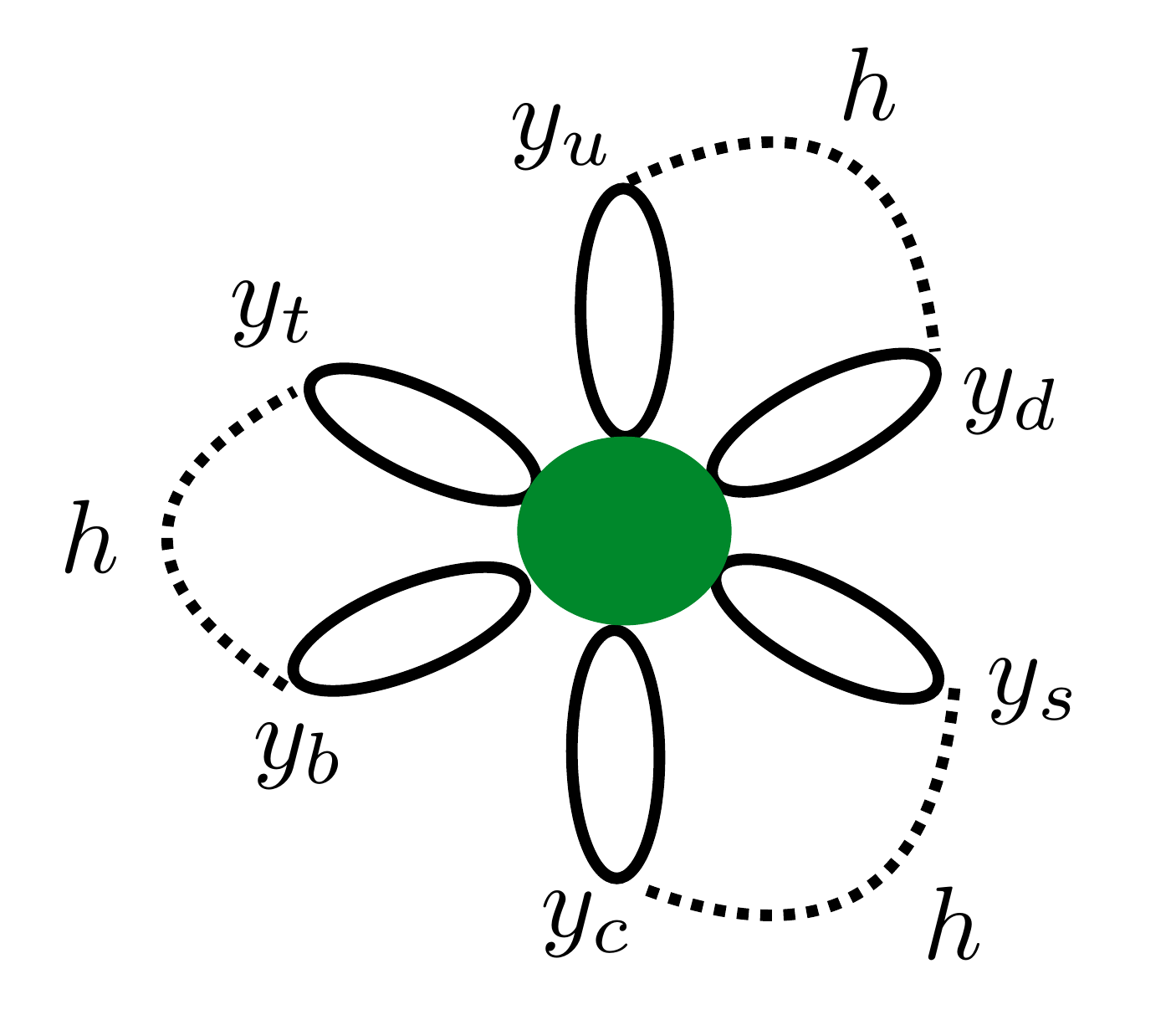}
\caption{An instanton vacuum-diagram schematically generates a short
  distance contribution to the axion potential that is proportional to
  the breakings of the individual quark chiral $U(1)$ factors by the Higgs
  Yukawa couplings. The dominant contribution at short distances is
  proportional to the Higgs vacuum fluctuations, corresponding to
  looping off the Higgs propagators.
  \label{fig:ChiralSuppressionInstanton} }
\end{figure}

Since the instanton contribution dominantly comes from the IR,
near the scale $M$, the instanton density in the
integrand \ref{eq:LambdaInt2} can be approximated as
\begin{align}
  D[\alpha_{s_{1,2}}(1/\rho)] 
  &\approx
  D[\alpha_{s_{1,2}}(M)]  
  (\rho M)^{b_{1,2}},
  \label{eq:DensityApprox}
\end{align}
where the one-loop running coupling satisfies
\begin{align}
  \frac{d\alpha_{s_i}^{-1}}{d\ln\mu} 
  &= \frac{b_i}{2\pi}
\end{align}
with $b_1 = 13/2$
and $b_2 = 21/2$ (including the contributions of the bifundamental
scalar $\Sigma$ to the running).  This gives
\begin{align}
  \Lambda_2^4 
  &=
  \frac{4}{13} D[\alpha_{s_2}(M)] M^4 
\end{align}
In the $SU(3)_1$ factor, there is a further suppression due to the
Yukawa couplings and Higgs loops, as depicted in
figure~\ref{fig:ChiralSuppressionInstanton}, which can be estimated as
\cite{Flynn:1987rs,Choi:1998ep}
\begin{align}
  \Lambda_1^4 
  &\sim 
  K \int_{\rho=0}^{\rho=1/M} 2 \frac{d\rho}{\rho^5} D[\alpha_{s_1}(1/\rho)] 
  \approx 
  K \frac{4}{5}  D[\alpha_{s_1}(M)] M^4
  ,
  \label{eq:Lambda1}
\end{align}
Where $K$ is a chiral suppression factor capturing the breaking of the $U(1)^6$ axial symmetry of the individual quarks by the Yukawa couplings\footnote{Ref.~\cite{Flynn:1987rs} considers the case where Yukawa couplings to a scalar state of mass $m_s$ break the chiral symmetries of the colored fermions in the theory. Their result contains an additional suppression factor of $m_s^2 / M^2$, which would arise in theories where a scalar mass-insertion is necessary to violate some of the chiral symmetries, e.g. a two-higgs-doublet model. In contrast, in our case there are non-decoupling effects even in the limit $m_s\rightarrow0$.},
\begin{align}
  K 
  &= 
  \left(\frac{y_u}{4\pi}\right)
  \left(\frac{y_d}{4\pi}\right)
  \left(\frac{y_c}{4\pi}\right)
  \left(\frac{y_s}{4\pi}\right)
  \left(\frac{y_t}{4\pi}\right)
  \left(\frac{y_b}{4\pi}\right) 
  \approx 10^{-23}.
\end{align}

We can compare the scales $\Lambda_{1,2}$ of the UV contributions to
the scale $\Lambda_0$ of the usual non-perturbative axion potential
generated around the QCD scale~\cite{Weinberg:1977ma} given in
equation~\eqref{eq:maNormal}.

\begin{figure}
\centering
\includegraphics[width=4in]{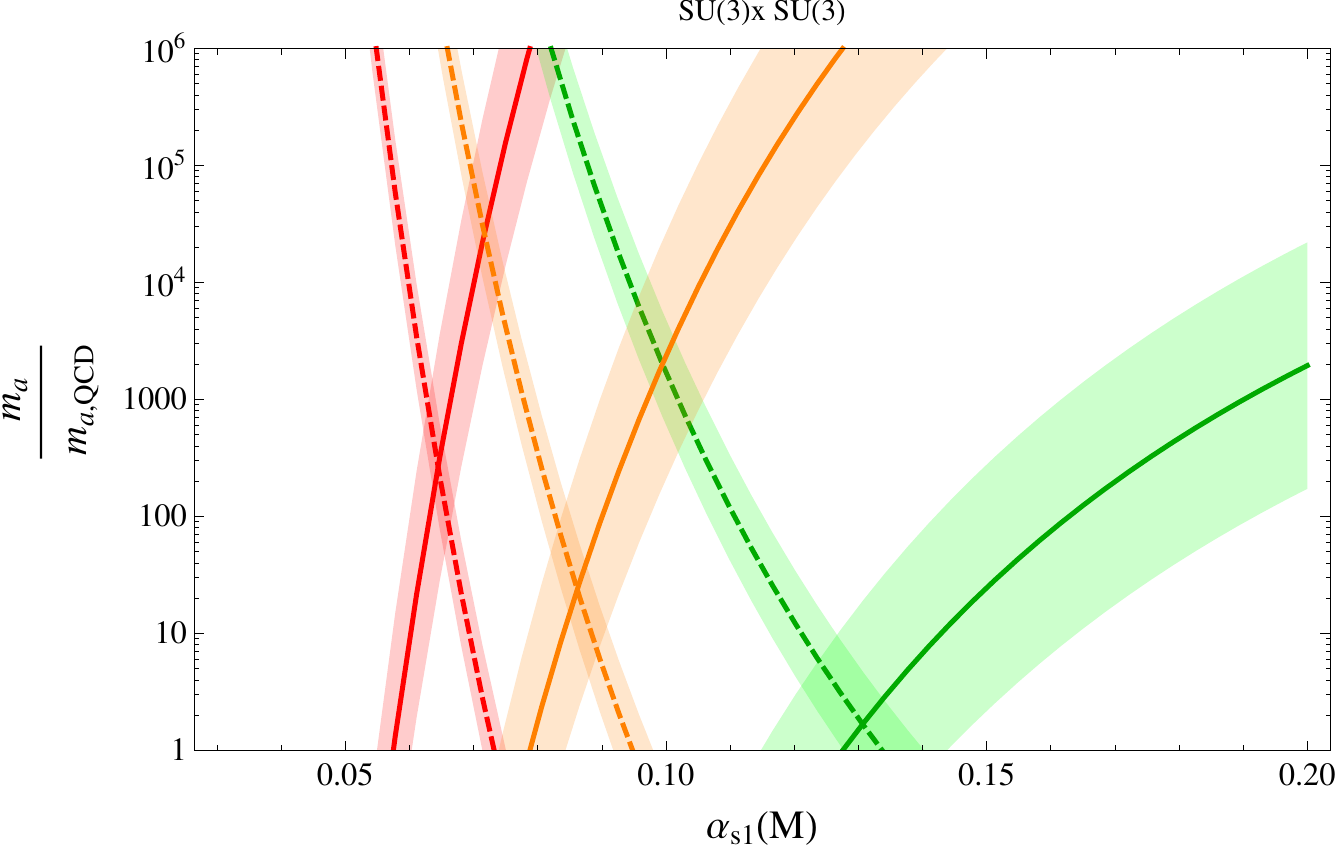}
\caption{The scale of the UV contribution to the axion mass as a
  function of $\alpha_{s_1}$, compared to the mass $m_{a,QCD}$ of a standard QCD axion given the same value of $f_a$. The solid lines give $m_{a_1}$ and
  the dashed lines give $m_{a_2}$, with $\alpha_{s_2}$ determined from $\alpha_{s_1}$
  by the matching condition equation~\eqref{eq:GaugeMatching}. The results
  are shown for a range of values of strong coupling scales, from left
  to right $M=10^{14}\GeV, 10^{11}\GeV, 10^{8}\GeV$  (red, orange, green). The parameter space of interest occurs when both $m_{a_1}/m_{a,QCD} \gg 1$ and $m_{a_2}/m_{a,QCD} \gg 1$ simultaneously, so that neither state has properties similar to the standard QCD axion.}
\label{fig:ScalesSU3xSU3}
\end{figure}

The size of the non-perturbative UV potentials generated in each
sector are shown in figure~\ref{fig:ScalesSU3xSU3}. The sensitivity to
the large scale $M$ can overcome the non-perturbative and
chiral suppression factors. The relation  equation~\eqref{eq:GaugeMatching}
between $\alpha_{s_1}$ and $\alpha_{s_2}$  implies that
$\Lambda_1$ grows as $\Lambda_2$ shrinks. In much of the parameter
range, $\Lambda_1 \gg \Lambda_0 \gg \Lambda_2$ or $\Lambda_2 \gg
\Lambda_0 \gg \Lambda_1$. In these cases, one axion DOF will simply
decouple, and the other will behave largely like the QCD axion. 

The most interesting regime is instead when $\Lambda_1, \Lambda_2  \gg
\Lambda_0$ -- in this case there will be no light state resembling the
standard QCD axion. This regime generally corresponds to
$\alpha_{s_1} > \alpha_{s_2}$ to compensate for the extra chiral
suppression factor in $\Lambda_1$. 
As the scale $M$ increases,  the scales $\Lambda_{1,2}$ grow --
although the running of $\alpha_3$ to weak coupling at higher energies
leads to stronger instanton suppression factors, this is overcome by
the $M^4$ dependence of the potential. For
$M \sim 10^8\GeV$, the UV effects can become
comparable in size to the IR potential, with $\Lambda_1\sim\Lambda_2
\sim \Lambda_0$ . For $M\sim 10^{14}\GeV$,  effects
as large as $\Lambda_1 \sim \Lambda_2 \sim 30 \Lambda_0$ can be
realized. This realizes a model with two axions, each $\sim 1000$
times heavier than a standard QCD axion with the same decay constant $f$, with a potential aligned to
dynamically set $\bar\theta=0$. Note that although the non-perturbative effects have a large effect on the axion masses in this regime, the individual gauge factors are still reasonably weakly coupled at the scale $M$, with $\alpha_{s_{1}}$ far away from the chiral-symmetry breaking phase which would be expected to occur at $\alpha_{s1} \gtrsim 0.7-1$ \cite{Roberts:1994dr,Appelquist:1997gq}.

So far we have considered the case $f_a \gtrsim M$ and assumed that
all states in the spontaneous PQ-breaking sector are decoupled. If
some of these states are lighter than $M$, the instanton effects can
be  suppressed. In a KSVZ axion model \cite{Kim:1979if,Shifman:1979if}
with $N_f$ flavors of vector-like quarks of mass $M_\psi \lesssim M$,
there will be a suppression in the instanton calculation by a factor
of $\sim (M_\Psi/\Lambda)^{N_f}$. We will take the benchmark $M_\Psi =
f$ and $N_f=1$ for the phenomenological studies in
Sec.~\ref{sec:pheno}, giving
\begin{equation}\label{eq:malowscale}
m^2_{a_i} \approx 
\begin{cases}
	\frac{\Lambda_i^4}{f_{a_i}^2} \quad f_{a_i} > M
        \vspace{1mm}
        \\
	\frac{\Lambda_i^3}{f_{a_i}} \quad f_{a_i} < M
\end{cases}.
\end{equation}
Other benchmarks like a KSVZ axion with $N_f > 1$ or a DFSZ axion \cite{Dine:1981rt,Zhitnitsky:1980tq} would typically have more severe suppressions for the regime $f_{a_i} < M$. 

\subsection{$SU(3)^N$ products}

\begin{figure}
\centering
\includegraphics[width=4in]{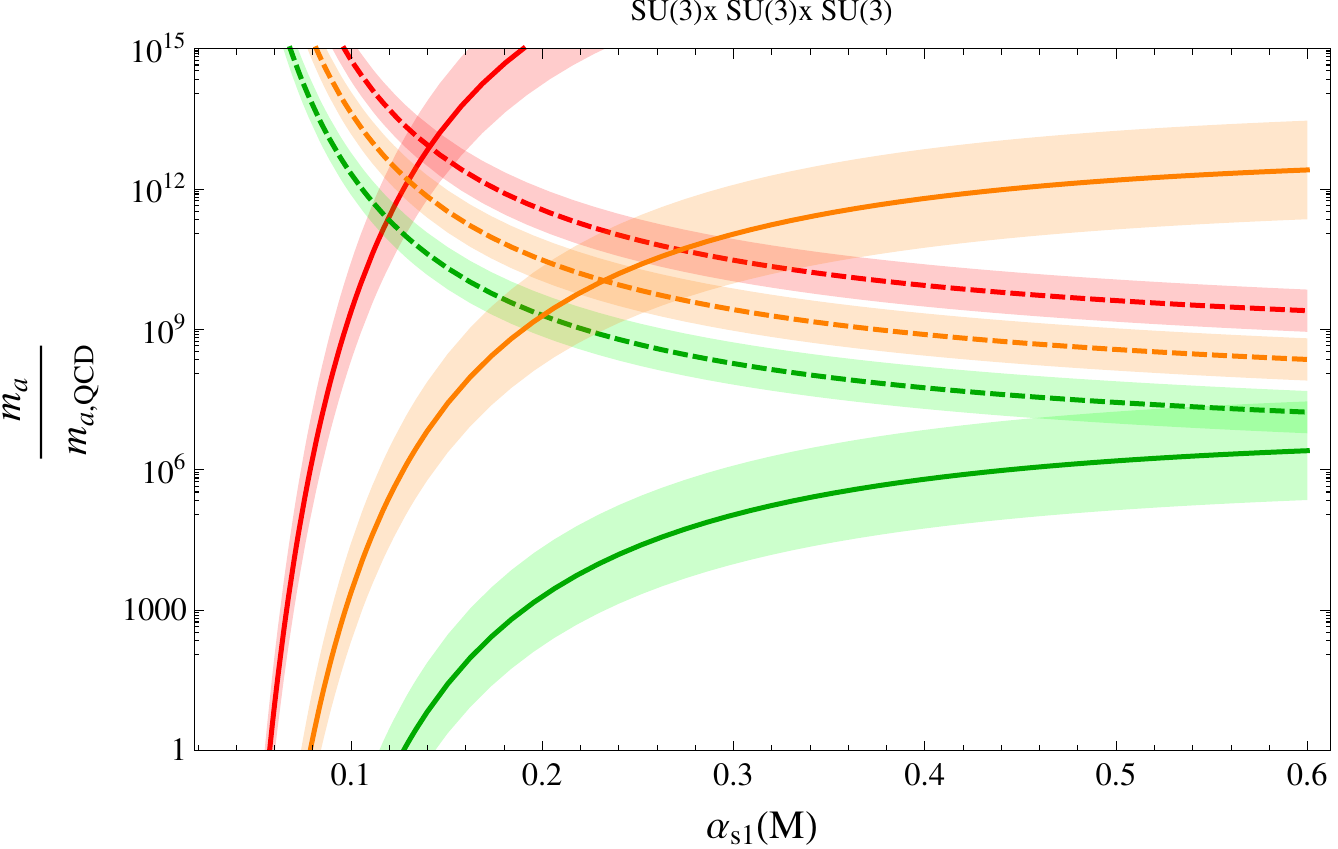}
\caption{The scale of the UV contribution to the axion mass as a
  function of $\alpha_{s_1}$, compared to the mass $m_{a,QCD}$ of a standard QCD axion given the same value of $f_a$. The solid lines give $m_{a_1}$ and
  the dashed lines give $m_{a_{2,3}}$, with $\alpha_{s_{2,3}}$ determined from $\alpha_{s_1}$ 
  by the matching condition equation~\eqref{eq:GaugeMatching} taking $\alpha_{s_2}=\alpha_{s_3}$. The results are shown for a range of
  values of strong coupling scales, from top to bottom $M=10^{14}\GeV, 10^{11}\GeV, 10^{8}\GeV$  (red, orange, green). The parameter space of interest occurs when both $m_{a_1}/m_{a,QCD} \gg 1$ and $m_{a_{2,3}}/m_{a,QCD} \gg 1$ simultaneously, so that neither state has properties similar to the standard QCD axion.}
\label{fig:ScalesSU3xSU3xSU3}
\end{figure}

Although each gauge factor in the $SU(3)\times SU(3)$ model is more
strongly coupled than the SM $SU(3)_{QCD}$, they remain relatively
weakly coupled, and the additional short distance non-perturbative
effects still suffer a significant exponential suppression. Extending
the gauge group to $SU(3)^N$, with the $SU(3)_{QCD}$ emerging from the
diagonal subgroup  allows the gauge coupling in each individual factor
to be substantially increased,
\begin{align}
  \frac{1}{\alpha_s}(\mu)
  &= \sum_{i=1}^{N} \frac{1}{\alpha_{s_i}(\mu)}
  \,,\qquad 
  \mu=M
  .
  \label{eq:GaugeMatchingN}
\end{align}
The gauge group can be Higgsed to the diagonal subgroup at a scale
$M$ by including multiple Higgs link fields $\Sigma_{12},
\ldots \Sigma_{N-1,N}$. An axion in each sector $a_1, ..., a_N$ can
remove each $\theta_1,..,\theta_N$. As for the $SU(3)\times SU(3)$
case, we take all of the SM quarks charged under $SU(3)_1$, and assume
no additional colored fermions are present in the other $SU(3)$ factors.
Even for $N=3$, the scale of the axion potential can now be
dramatically increased compared to the standard QCD axion.
Figure~\ref{fig:ScalesSU3xSU3xSU3} shows the $N=3$ case taking
$\alpha_{s_2}=\alpha_{s_3}$. For  example for
$M=10^{14}~\GeV$, a model can be realized with 3 axions
all with masses $\sim 10^{12}$ times larger than a standard QCD
axion with the same decay constant $f$, again with a potential aligned to dynamically set
$\bar\theta=0$.

It is interesting to ask how large the axion mass can be made if an even larger number of product factors are introduced. For $N\sim 10$, the couplings of the individual SU(3) factors can be
made sufficiently strong to generate $\mathcal O(1)$
non-perturbative effects, giving $N-1$ axions with masses $m_{a_i}
\sim {M}^2 / f_i$, while the mass of the axion in the $SU(3)_1$ sector
will still be suppressed by the chiral suppression factor, giving $m_{a_1}\sim
10^{-12} {M}^2/f_1$. 

\section{UV Sensitivity}
\label{sec:UV}

There are several interesting questions about the sensitivity of this
multi-axion solution to UV physics above the scale $M$.  First, as is
familiar from the standard QCD axion picture, Planck-suppressed
operators may spoil some of the PQ symmetries
\cite{Barr:1992qq,Kamionkowski:1992mf,Ghigna:1992iv}, naively leading
to a misalignment of the axion potential by 
\begin{align}\label{eq:PQquality}
\Delta \bar\theta \sim
\frac{f_a^{(d-2)}}{M_{pl}^{(d-4)} m_a^2}, 
\end{align}
with the dimension $d \geq 5$ of the operator depending on
the details of the axion model.  In our case, this problem is
significantly alleviated compared to the standard QCD axion because
smaller values of $f_a$ and larger values of $m_a$ can be realized.
Fig.~\ref{fig:mafa} shows that large parts of parameter space remain
open even assuming the most dangerous $d=5$ operators are present.

In our model there are also PQ-preserving higher dimensional operators that can be dangerous. Focusing for simplicity on the $SU(3)\times SU(3)$ case, at $D=6$ the operator 
\begin{equation}
\frac{\Sigma_{12}\tilde G_2 \Sigma_{12}^\dagger G_1}{\Lambda_{UV}^2}
\end{equation}
 can introduce a shift $\Delta \bar  \theta \sim {M}^2 / \Lambda^2_{UV}$ that is not canceled at the axion minima. 
Requiring $\bar\theta \lesssim 10^{-10}$ then requires a hierarchy $M
\lesssim (10^{14}~\GeV) \left(\frac{\Lambda_{UV}}{M_{pl}}\right)^2$.

Another possibility is that non-perturbative effects may not
decouple sufficiently rapidly to protect the theory from CP violation at scales far above $M$. For example, if there are new unsuppressed sources of CP violation in the theory at the scale
$\Lambda_{UV}$, then non-perturbative effects at this scale can
generate additional misaligned contributions to the axion potential.
Because each gauge factor is asymptotically free, the non-perturbative
effects decouple at high energies as $\alpha_{s_i}$ run to weak
coupling. The $SU(3)_1$ factor contains the SM quarks, and therefore
runs to weak coupling the most slowly. From
equations~\eqref{eq:Lambda1},\eqref{eq:DensityApprox}, the scale of the
potential generated for $a_1$ at $\Lambda_{UV}$ is suppressed compared
to the potential generated at $M$ by a factor
\begin{equation}
\left(\frac{M}{\Lambda_{UV}}\right)^{b_1-4} =\left(\frac{M}{\Lambda_{UV}}\right)^{5/2}
\end{equation}
Assuming an $\mathcal{O}(1)$ misalignment of the potential generated
at $\Lambda_{UV}$, the limit  $\bar\theta \lesssim 10^{-10}$ requires
a hierarchy $M\lesssim (10^{15}\GeV)
\left(\frac{\Lambda_{UV}}{M_{pl}}\right)^{5/2}$.

Therefore without additional assumptions about the nature of CP
violation at the scale $\Lambda_{UV}$, there must be a substantial
hierarchy in scales $M\ll \Lambda_{UV}$ to protect against higher dimensional operators and sufficiently suppress non-perturbative effects. In the simple model we have presented,
$M$ is set by an elementary scalar field $\Sigma_{12}$
breaking the gauge group, and it seems that maintaining this hierarchy
reintroduces as severe of a tuning as the tuning in $\bar\theta$ we
have set out to address. Two well-known solutions exist -- this
hierarchy can be protected by supersymmetry (SUSY) or by a
technicolor-like mechanism. 

In a SUSY model, the scale $M$ can be protected if
$m_{\rm soft} \lesssim M$. When $m_{\rm soft}\ll$
$M$, the instanton-generated potential will be suppressed
by  a factor of at least $(m_{\rm soft}/M)^5$ due to
insertions of gaugino masses and the PQ-breaking $b H_u H_d$ soft
terms \cite{Choi:1998ep}, so it is desirable to stay in the regime
$m_{\rm soft}\sim M$. Since $m_{\rm soft}\sim
M\gg 100~\GeV$, it is not possible to protect the weak
scale with SUSY in this model, but at least the hierarchy
$M \ll \Lambda_{UV}$ can be maintained. However, the soft
SUSY breaking introduces new CP phases into the theory, and the
alignment of the axion potential which gives $\bar\theta=0$ is no
longer guaranteed. As is familiar from low-scale SUSY, we must assume
that the soft terms are communicated in a way that does not introduce
new CP violating phases (for a review, see Ref.~\cite{Martin:1997ns}), since these would lead radiatively to misalignment of the short-distance axion potential generated near
$M$ (the standard QCD axion mechanism is similarly
sensitive to the CP violation in the soft sector when $m_{\rm soft}
\lesssim 100 \TeV$ \cite{Hamzaoui:1998yu}). Ref.~\cite{Dine:2015jga}
points out a similar sensitivity to SUSY-breaking CP violation in
Nelson-Barr models. 

In a technicolor-like solution, the $\Sigma$ can be made a composite
of elementary fermions under a new strong group, with the scale $M$ generated dynamically. This resembles asymptotically free completions of the dimensional deconstruction framework \cite{ArkaniHamed:2001ca}. However, because new fermions charged under $SU(3)_1$ and $SU(3)_2$ will be introduced, we must take care that their additional chiral symmetries are broken in a way that does not suppress the non-perturbative axion potential or re-introduce new CP phases. An example of such a model is given in Appendix~\ref{app:composite}.

\section{Phenomenology}
\label{sec:pheno}
Laboratory experiments, cosmology, astrophysics, beam dump, and
collider experiments can all be sensitive probes of new light
pseudoscalars, also known as axion-like particles. Their reach can be
parameterized in the space of masses $m_a$ and the dimension-5
couplings of the new states, which scale as $\frac{1}{f_a}$. Although
our models contain multiple axion-like states, the direct couplings
in-between these states are phenomenologically unimportant and we can
treat the limits on each axion-like state independently.

Since we have not specified the UV details of the spontaneous PQ-breaking,
the heavy axions in our models may realize dimension-5 couplings to
the electroweak gauge bosons  $a W\tilde W, a B\tilde B,$ and light
fermions $\partial^\mu a (f^\dagger \gamma_5 \gamma_\mu  f)$ in
addition to the gluon coupling $a \tilde G G$. When $m_a < m_{\eta'}
\sim \GeV$, these additional couplings to the mass eigenstates are
inevitable at low energies since the axions mix with the $\eta'$,
$\eta$, and $\pi_0$ mesons \cite{Fukuda:2015ana}, as in
the standard QCD axion model
\cite{Weinberg:1977ma,Wilczek:1977pj,diCortona:2015ldu}. When the
axion is light enough that hadronic decay channels are closed, $m_a
\lesssim 3 m_\pi$, the phenomenology is primarily determined by the
low-energy axion-photon coupling, and this ALP case has been well
studied.  

For $m_a \gtrsim ~1 \GeV$, hadronic decay channels open and typically
dominate, and studies focused on an axion decaying dominantly to
electroweak gauge bosons no longer apply.  Limits  depending on these
couplings are even further weakened when UV couplings to the
electroweak gauge bosons are absent, since the mixing-induced
couplings become rapidly suppressed as QCD runs to weak coupling. 

To study the limits on our models, we take as a benchmark an effective
low-energy photon coupling of 
\begin{align}
  \mathcal{L}_{a\gamma\gamma}
  &=
  \frac{e^2}{32\pi^2}\frac{a_i}{f_i} F
  \tilde F 
  \label{eq:photonCoupling}
\end{align}
throughout the parameter space
(for $m_a \gg m_{\eta'}$, the photon coupling will be entirely due to
the UV physics, while for $m_a \lesssim m_{\eta'}$ it is a result of
both the UV coupling and the mixing with the neutral mesons). We use
this simplified approximation as a rough guide to the phenomenology
and a more realistic treatment would generically include couplings to
the $Z$ as well as to other SM particles.

\begin{figure}[t]
\centering
\includegraphics[width=0.9\textwidth]{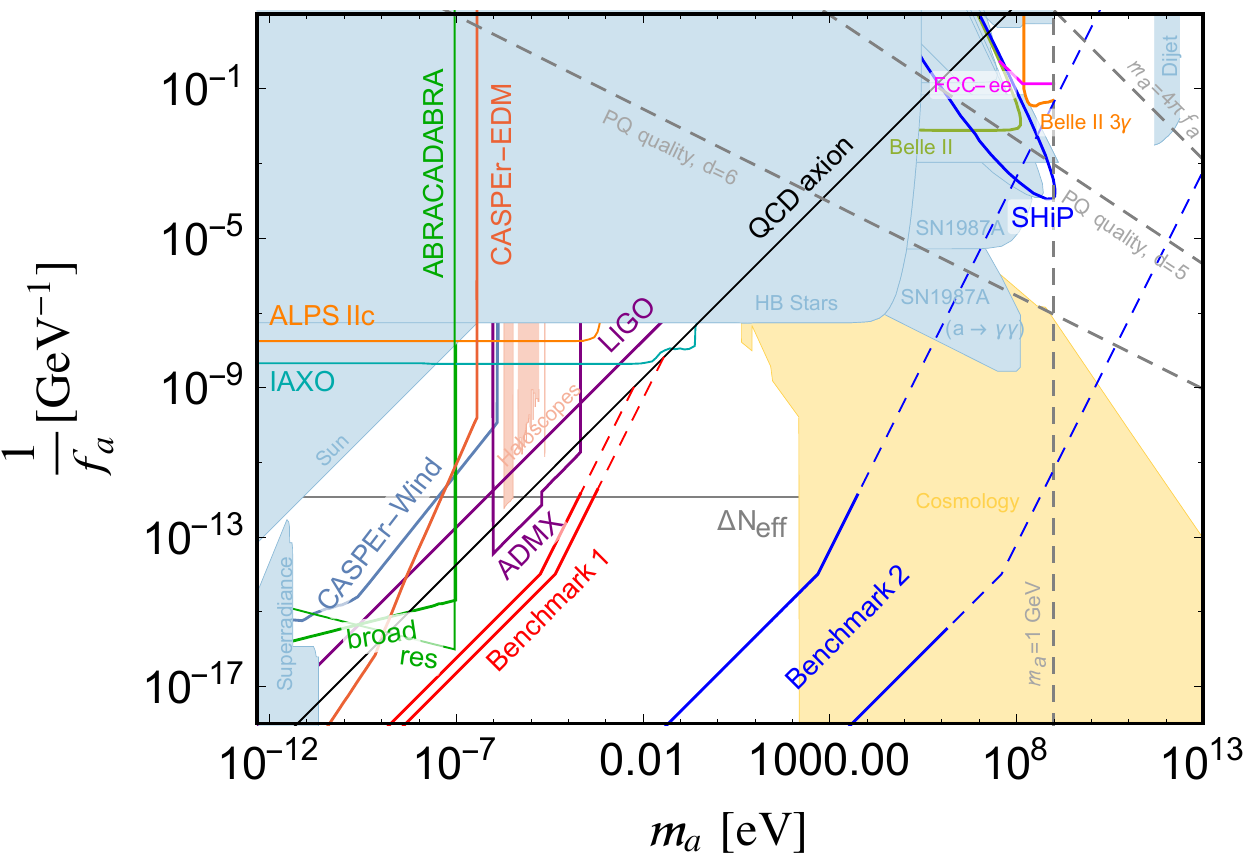}
\caption{\label{fig:mafa}
Limits on axions in the $(m_a, f_a)$ plane for an axion with couplings
to gluons and photons given by  equations~\eqref{eq:LUV} and
~\eqref{eq:photonCoupling}. The black solid line is the standard QCD
axion, with mass given by equation~\eqref{eq:maNormal}. Parallel to this
line from left to right are the lines for  $a_1$ and
$a_2$ (red) in the $SU(3)\times SU(3)$ benchmark parameters
described in the text, and  $a_{2,3}$ and $a_1$ (blue) for the
$SU(3)\times SU(3) \times SU(3)$ benchmark parameters.
The solid segments on these lines indicate where they could be
realized as dark matter from the misalignment mechanism.
The shaded regions show current constraints, and the regions bounded
by solid lines are sensitivities of future experiments (see text for
details).
}
\end{figure}

We show existing
limits and future projections for axion experiments as a function of
$m_a,f_a$ in
figure~\ref{fig:mafa}. We also 
include
two benchmark models in which the axions are lifted to
masses much larger than the standard QCD axion. 

The first benchmark model is the $SU(3)\times SU(3)$ model with
$\alpha_{s_1}(M) = 0.063$ and $M=10^{14}\GeV$. This
model contains two axions, each corresponding to a separate curve in
the $(m_a, f_a)$ plane. In principle, $f_1$
and $f_2$ can occur at very disparate scales, so each axion could
appear at any point on its curve, although it might be appealing to
connect the scales $f_1$ and $f_2$ in a detailed model of the
spontaneous PQ-breaking sector. At small values of $f_a$, both
$\Lambda_1$ and $\Lambda_2$ become less than $\Lambda_0$. In
this region one linear combination behaves as the usual QCD axion,
and the orthogonal combination is light and has suppressed couplings
to QCD.

The second benchmark model is the $SU(3)\times SU(3) \times SU(3)$
model with $\alpha_{s_1}(M) = 0.1$,
$\alpha_{s_2}=\alpha_{s_3}$, and $M=10^{14}\GeV$.
Of particular interest is a new region
in parameter space for a $f_a\sim \TeV$ weak-scale axion that appears
to be open in this model when $m_a \gtrsim \GeV$ -- it would be
interesting to study in further detail the limits on such a
hadronically decaying state.

We have chosen these two specific benchmarks to illustrate the
parameter space and the correlations between the scales of the
multiple axions present in each model. We note however, that the
parameter space of these models motivates axions spanning the entire
$(m_a, f_a)$ plane above the standard QCD axion line.

We briefly describe the constraints and projections shown in
figure~\ref{fig:mafa}.  Current constraints are shown as shaded
regions. The light blue shaded region are independent of a dark matter
interpretation of the axion or its cosmology. The yellow shaded
regions assume that the reheating temperature was large enough to have
thermally produced axions, and that there was no large entropy dump to
the SM after axion decoupling.  The red shaded regions assume that the axion
makes up all of the dark matter.  

{\bf Misalignment and $\Omega_{\rm{DM}}$}:
In parts of our parameter space, axions can make up all
of the dark matter. In our model the mass of the axion goes to its
zero temperature value much earlier than the time it starts
oscillating, making the computation of relic abundance much simpler
than the standard QCD axion case. In our case the relic abundance for
a single axion with mass $m_a$ and decay constant $f_a$ is
given as,
\begin{align}
  \Omega_a h^2
  &=
  \frac{m_a^2 f_a^2 \theta_0^2}{\rho_{\rm{crit}}/h^2}
  \left(\frac{a(T_f)}{a_0}\right)^3 
  \simeq
  0.12
  \left(\frac{f_a}{10^{12}\GeV}\right)^2
  \left(\frac{m_a}{0.01\eV}\right)^{1/2}
  \theta_0^2
\end{align}
where $a(T_f)$ is the scale factor at the temperature $T_f$ where the axion starts oscillating,
$H(T_f)\simeq m_a/3$, $\theta_0$ is the initial misalignment angle,
and $\rho_{\rm{crit}} = 3.96\times10^{-47} \GeV^4$ is the critical
density of the Universe. This estimate ignores the thermal production
of axions, which can become important for higher masses as well. We
show the part of the parameter space in our benchmarks where heavy
axions could make up the dark matter by solid lines in
figure~\ref{fig:mafa}. It is interesting to note that in our models
there is a new region of the $m_a$--$f_a$ parameter space where axions
solve the strong CP problem and constitute all of the dark matter for
generic initial conditions ($\theta_0 \sim 1$).

{\bf Cosmology constraints}
The yellow region is ruled out by a combination of cosmological
constraints, assuming that the reheat temperature was sufficiently
high to produce an equilibrium thermal population of axions
\cite{Cadamuro:2011fd} which dominates over misalignment production in
this region~\cite{Arias:2012az}.
Future CMB observations can also put significant constraints on
thermally produced axions from measurements of $\Delta N_{\rm{eff}}$ that
constrains extra relativistic degrees of freedom present during
recombination. We show the sensitivity of the CMB-S4 assuming a reheat
temperature of $T_R=10^{10}\GeV$~\cite{Baumann:2016wac} with a
horizontal gray line.

If the axion makes up dark matter, then its early universe values can
be large enough to affect BBN~\cite{Blum:2014vsa}. This bound is
superseded by the solar bound mentioned below.

{\bf Astrophysical bounds}
Some of the strongest constraints on the axion parameter space arise
from stellar physics, where axion coupling to photons is constrained
by evolution of horizontal branch stars. For lower masses of axions
supernova constraints and constraints from
conversion of X-ray photons to axions in cluster magnetic fields
\cite{Berg:2016ese,Marsh:2017yvc,Conlon:2017qcw} impose a stronger bound.
Axion helioscopes such as
CAST~\cite{Anastassopoulos:2017ftl} also put constraints on the couplings
of axions with photons, with the future IAXO
helioscope~\cite{Vogel:2013bta} improving the reach.
The light blue region includes these
astrophysical constraints (adapted from
\cite{Jaeckel:2015jla}). 

Axions which lie above the QCD line in figure~\ref{fig:mafa} need a
tuned negative contribution to their mass to cancel the partially
cancel the mass from QCD. Their potential can then flip its sign due
to finite density effects in astrophysical objects, which then source
the axion field.  This observation was used by
Ref.~\cite{Hook:2017psm} to put bounds on this parameter space by
dense astrophysical objects sourcing axion fields. In
figure~\ref{fig:mafa} we show the current bounds arising from the Sun
and projections from neutron star mergers in LIGO (purple), which may
be able to probe most of this tuned region of parameter space. We use
$m_a = 0.1 m_{a,QCD}$ as a conservative projection from LIGO.  Black
hole superradiance~\cite{Arvanitaki:2014wva} puts constraints on very
light axions.

{\bf Collider constraints:}
The possibility of heavier QCD axions also opens up weak-scale decay
constants, making these axions visible at colliders. 
We show constraints from monojet searches at the
LHC \cite{Mimasu:2014nea} and 
from LHC dijet \cite{Aaboud:2017yvp,CMS-PAS-EXO-16-056} and diphoton
searches \cite{ATLAS-CONF-2016-059,Khachatryan:2016yec} for
pseudoscalar states produced in gluon fusion~\cite{Belyaev:2016ftv}.
Also show are the constraints from by monophoton and beam dump
experiments (adapted from \cite{Dolan:2017osp}).  
The Belle II experiment~\cite{Abe:2010gxa} will be
sensitive to heavier axion-like
particles at the weak scale~\cite{Dolan:2017osp}. The SHiP beam dump
experiment~\cite{Anelli:2015pba} can
cover intermediate mass scale axions which are not
yet excluded by current experiments. Future $e^+e^-$ colliders on the
$Z$ mass peak can also improve on axion couplings. We show
projections~\cite{Jaeckel:2015jla} for the FCC-ee
collider~\cite{Gomez-Ceballos:2013zzn}.

Some of the analyses above were carried out for axion-like particles
which only have a coupling to photons, and hence do not directly apply
to our case when decays to hadrons become important. 
As a crude fix we have cut off all such limits at $m_a = 1 \GeV$. It
would be interesting to recast collider bounds on models of heavy QCD
axions including couplings to gluons as well as photons and other SM
particles.

{\bf Laboratory experiments:}
Laboratory experiments such as the haloscope ADMX constrains the QCD
axion in the region where it can be the dark matter.
We show regions of parameter space that can be covered by future
experiments. 
The upgraded cavity experiment ADMX2~\cite{Shokair:2014rna} will cover a larger
range of QCD axion masses with higher sensitivity. The CASPEr experiments~\cite{Graham:2013gfa,Budker:2013hfa,Garcon:2017ixh}
propose using NMR
techniques to measure time-varying EDMs induced by the axion, which
will be sensitive for low-mass axions. The ABRACADABRA~\cite{Kahn:2016aff}
experiment can also cover the low mass axion parameter space for
axion-photon couplings. These
experiments all rely on the axion being all of dark matter.
The upgraded light-shining-through-wall experiment ALPS
II~\cite{Bahre:2013ywa} will be sensitive to larger couplings
of axion-like particles
with the photon.

We have included in figure~\ref{fig:mafa} some dashed gray lines motivated
by theoretical
considerations. The line in the upper right corner signals the invalidity of
the effective axion theory when $m_a \gtrsim 4\pi f_a$. The reduced
sensitivity to quality of the PQ symmetry through Planck suppressed
higher dimensional operators as described by equation~\eqref{eq:PQquality} are also shown. In regions above these lines, generic dimension--5
and --6 operators do not spoil the solution to the strong CP problem.
The vertical dashed gray line at $m_a =1~\GeV$ shows the scale where 
hadronic decays begin to dominate
and axion mixings with the neutral mesons begin to be suppressed.

A few general comments which follow from our survey of the
phenomenology:
\begin{itemize}
  \item 
In our plot, most of the current and future observations only probe
the region above the QCD line. As noted above, this region is tuned
such that there is a negative mass contribution to the axion, with a
minimum which is highly aligned with the QCD minimum, and a size
comparable to the QCD contribution.  However, experiments which proble
the axion coupling to photons can extend below the QCD line in our
plot if the coupling of the axion to photons is much larger than the
coupling assumed here (as
in~\cite{DiLuzio:2016sbl,Farina:2016tgd,Agrawal:2017cmd}). With such
an enhanced couplings these experiments will also be sensitive to
heavier QCD axions as considered in this paper.

\item When hadronic channels are open, the branching ratio to photons
  is tiny. The collider signals of a heavy QCD axion therefore differ
  from the more-often studied case of ALPs which only have a photon
  coupling. Our work shows that the combination of gluon and photon
  couplings is very well motivated even for heavy collider-observable
  axions and deserve further study. 
\item 
  We also note that the phenomenology of this realization of
  a heavy axion is substantially different from the
  phenomenology in a $Z_2$ heavy axion model, where the axion
  may decay to light mirror-sector particles
  \cite{Fukuda:2015ana}, and is generically not expected to appear
  with
  multiple copies.
\item A combination of heavy axions can make up the dark matter
  density through the
  misalignment mechanism in a new part of $m_a$--$f_a$ plane,
  motivating searches in this region. A resonance search strategy for
  laboratory experiments
  might be less optimal if the dark matter density is not dominantly
  stored in axions at one mass.
\end{itemize}

\section{Conclusions}

We have described a novel mechanism for solving the strong CP problem.
In the standard QCD axion mechanism, a spontaneously broken anomalous
$U(1)_{PQ}$ symmetry results in an axion with a potential generated by
non-perturbative effects near the scale $\Lambda_{QCD}$. In the
absence of any UV sources of perturbative PQ violation, the strong CP
problem is solved dynamically when the axion relaxes to its CP
preserving minimum. In our mechanism, we embed QCD in a $SU(3)^N$
product group at a scale $M\gg\Lambda_{QCD}$, and use a separate
spontaneously broken PQ symmetry in each individual SU(3) factor to
dynamically relax each individual $\bar\theta$ angle. Because each
individual SU(3) factor is more strongly coupled than the SM QCD at
the scale $M$, the non-perturbative contributions to the axion
potentials at the scale $M$ can be much larger than those generated
near $\Lambda_{QCD}$ for the standard QCD axion.  After integrating
out physics above the scale $M$, the theory is just the standard model
with $N$ axions, each with a PQ-violating potential that arose from
non-perturbative effects near the scale $M$. Although the PQ
symmetries are explicitly violated in this low-energy effective
theory, the non-perturbative origin of the axion potentials in the
full theory guarantees that the minimum relaxes the low-energy
$\bar\theta$ angle while generating a mass for each axion that is much
larger than the standard QCD axion mass relationship.

Our assumptions about the existence of spontaneously broken anomalous
PQ symmetries are on the same footing as the standard QCD axion, and
it is encouraging that in general a large number of axions are
expected in string theory models \cite{Arvanitaki:2009fg}. In fact, in
our model, the possibility of solving the strong CP problem with a
small value of $f_a$ and large value of $m_a$ can resolve the problem
of maintaining a sufficient PQ quality in the presence of
quantum-gravity corrections that plague the standard axion solution
\cite{Barr:1992qq,Kamionkowski:1992mf,Ghigna:1992iv}.
 
The main obstacle to the model we have proposed is the conflict with
solutions to the electroweak hierarchy problem. Solving the
electroweak hierarchy problem in this class of model suggests $M\sim
\TeV$, while generating large axion masses requires $M \gg \TeV$.
While we have shown that it is possible to stabilize at least the
hierarchy $M \ll M_{pl}$, it requires further model building to ensure
that additional sources of CP violation do not spoil the mechanism. 

Phenomenologically, the model motivates exploring the whole $(m_a,
f_a)$ range of axion-like particles, including the range $m_a \gtrsim
3 m_\pi$ where hadronic decays will dominate. While the model requires
the existence of multiple heavy axion states, it would be difficult to
directly verify their connection to the strong CP problem, as the
precision mass-coupling relationships predicted for the standard
QCD-axion are no longer realized. Because there is the possibility of
decoupling the heavy axions to scales $m_a \gg M_W$, this work can
also be viewed as an interesting new theoretical example of a
completely decoupling solution to the strong CP problem.

We would like to thank Asimina Arvanitaki, Savas Dimopoulos, Bogdan
Dobrescu, Roni Harnik, Anson Hook, Junwu (Curly) Huang, Gustavo Marques
Tavares, Lisa Randall and Raman Sundrum for encouragement, helpful
conversations, and comments on the manuscript.  This work was
initiated under a street-lamp at the Aspen Center for Physics, which
is supported by National Science Foundation grant PHY-1066293.  PA is
supported by the NSF grants PHY-0855591 and PHY-1216270.  This
manuscript has been authored by Fermi Research Alliance, LLC under
Contract No.  DE-AC02-07CH11359 with the U.S. Department of Energy,
Office of Science, Office of High Energy Physics. The United States
Government retains and the publisher, by accepting the article for
publication, acknowledges that the United States Government retains a
non-exclusive, paid-up, irrevocable, world-wide license to publish or
reproduce the published form of this manuscript, or allow others to do
so, for United States Government purposes.

\appendix
\section{Composite Link Fields}
\label{app:composite}

\begin{figure}
\centering
\includegraphics[width=2in]{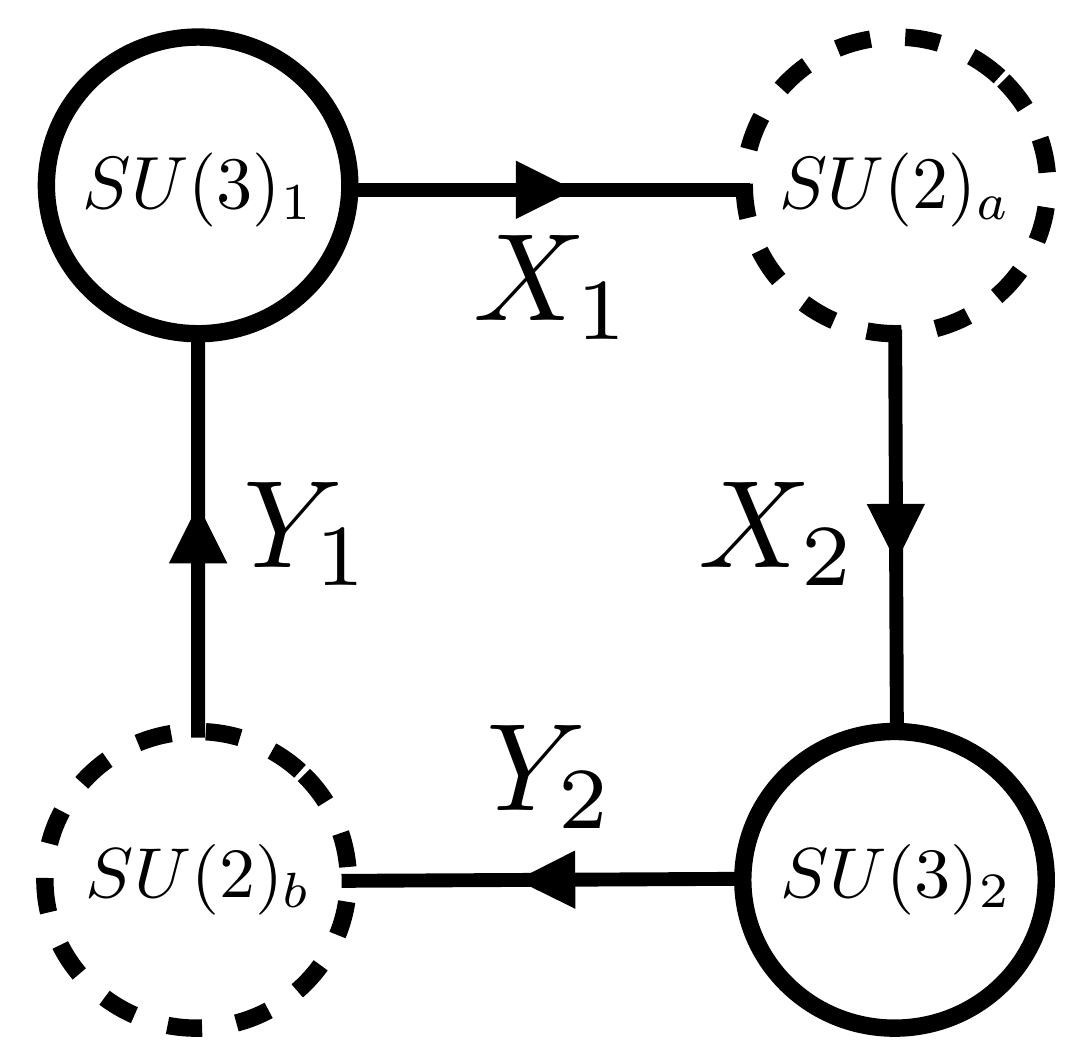}
\caption{\label{fig:Quiver}A moose model giving composite link fields
to stabilize the hierarchy $M \ll \Lambda_{UV}$.}
\label{fig:ScalesSU3xSU3xSU32}
\end{figure}

A model with composite link fields can stabilize the hierarchy
$M \ll \Lambda_{UV}$. 
For example, the $SU(3)_1\times SU(3)_2$ model can be extended to a
$SU(3)_1 \times SU(2)_a \times SU(3)_2 \times SU(2)_b$ moose theory as
shown in figure~\ref{fig:Quiver}, with condensation of elementary
fermion bilinears $\langle X_1 X_2 \rangle$ and $\langle Y_{1} Y_2
\rangle$ in the $SU(2)$ factors breaking $SU(3)_1 \times SU(3)_2
\rightarrow SU(3)_c$, making the scale $M$ dynamical.

In addition to the new gauge interactions,  gauge invariant 4-fermion
interactions of the form 
\begin{align}
  \mathcal{L}_{\slashed{\chi}}
  &=
  \frac{\lambda_1}{M} (X_1 X_1) (Y_1 Y_1)
  +\frac{\lambda_2}{M} (X_2 X_2)(Y_2 Y_2) 
\end{align}
are necessary to break the additional chiral symmetries of
the model. The instantons in $SU(3)_1$ and $SU(3)_2$ sectors will be
suppressed by the coefficients $\lambda_1$ and $\lambda_2$, so this
operator must either be generated at a nearby scale (without generating
additional dangerous CP violating operators or a new hierarchy
problem), or the theory must enter a conformal regime in the UV where
these 4-fermion operators have large negative anomalous dimension.

Field redefinitions can be used to choose $\lambda_{1,2}$ real, and
one remaining anomalous U(1) can be used to remove one combination of
the  $\theta$-angles in the $SU(2)_a$ and $SU(2)_b$ sectors. The
remaining new $\theta$ angle may be removed by coupling an additional
axion degree of freedom to the $SU(2)_a$ or $SU(2)_b$ sector, which
will obtain a large mass $\sim \lambda_{1}\lambda_2 {M}^2
/ f$.

\bibliographystyle{utphys}
\bibliography{HeavyAxionsRefs}

\end{document}